%% file: ijcai18.tex
\documentclass{article}
\usepackage{ijcai18}
\usepackage{times}
\usepackage{xcolor}
\usepackage{soul}
\usepackage[utf8]{inputenc}
\usepackage[small]{caption}
\usepackage{graphicx}
\usepackage{amsfonts, amssymb, amsmath}
\usepackage{multirow}
\usepackage{paralist}
\usepackage{rotating}
\usepackage{subfigure}
\usepackage{amssymb,bm,mathrsfs}
\usepackage{booktabs}
\newcommand{\newcite}[1]{\citeauthor{#1}~\shortcite{#1}}

\newcommand{\ie}{\textit{i.e.},}
\newcommand{\eg}{\textit{e.g.},}

\newcommand{\jumper}{\textsc{Jumper}}
\newcommand{\jumperb}{\textsc{Jumper }}

\newcommand{\gru}{\operatorname{GRU}}

\title{\textsc{Jumper}: Learning When to Make Classification Decisions in Reading}

\author{
Xianggen Liu$^{1}$\thanks{The work was done when the first author was as an intern at DeeplyCurious.ai},
Lili Mou$^2$, 
Haotian Cui$^{1}$,
Zhengdong Lu$^3$, 
Sen Song$^{1,4}$
\\ 
$^1$Department of Biomedical Engineering, \\IDG/McGovern Institute for Brain Research, Tsinghua University\\
 $^2$AdeptMind.ai\\
 $^3$DeeplyCurious.ai\\
 $^4$Laboratory of Brain and Intelligence, Tsinghua University\\
liuxg16@mails.tsinghua.edu.cn, doublepower.mou@gmail.com,\\ cht15@mails.tsinghua.edu.cn, luz@deeplycurious.ai, songsen@mail.tsinghua.edu.cn
}

\begin{document}
\maketitle

\begin{abstract}
 In early years, text classification is typically accomplished by feature-based machine learning models; recently, deep neural networks, as a powerful learning machine, make it possible to work with raw input as the text stands. However, exiting end-to-end neural networks lack explicit interpretation of the prediction. In this paper, we propose a novel framework, \jumper, inspired by the cognitive process of text reading, that models text classification as a sequential decision process. Basically, \jumperb is a neural system that scans a piece of text sequentially and makes classification decisions at the time it wishes. Both the classification result and when to make the classification are part of the decision process, which is controlled by a policy network and trained with reinforcement learning. Experimental results show that a properly trained \jumperb has  the following properties: (1) It can make decisions whenever the evidence is enough,  therefore reducing total text reading by 30–40\% and often finding the key rationale of prediction. (2) It achieves classification accuracy better than or comparable to state-of-the-art models in several benchmark and industrial datasets.
\end{abstract}

\section{Introduction}
Text understanding is one of the core goals of natural language processing (NLP), and is related to various applications, including text classification~\cite{Kim14sent}, information extraction~\cite{relation}, and machine comprehension~\cite{squad}. Recently, neural networks are playing an increasingly important role in NLP and have achieved significant performance in these tasks. However, previous work mainly focuses on the ultimate performance of a task (\eg\ classification accuracy). Humans typically do not have a clear understanding on where and how the model makes such a decision, which are in fact important for debuggability and interpretability especially in real industrial applications~\cite{criticism}. 
\begin{figure}[t]
	\centering
	\includegraphics[width=1.03\linewidth]{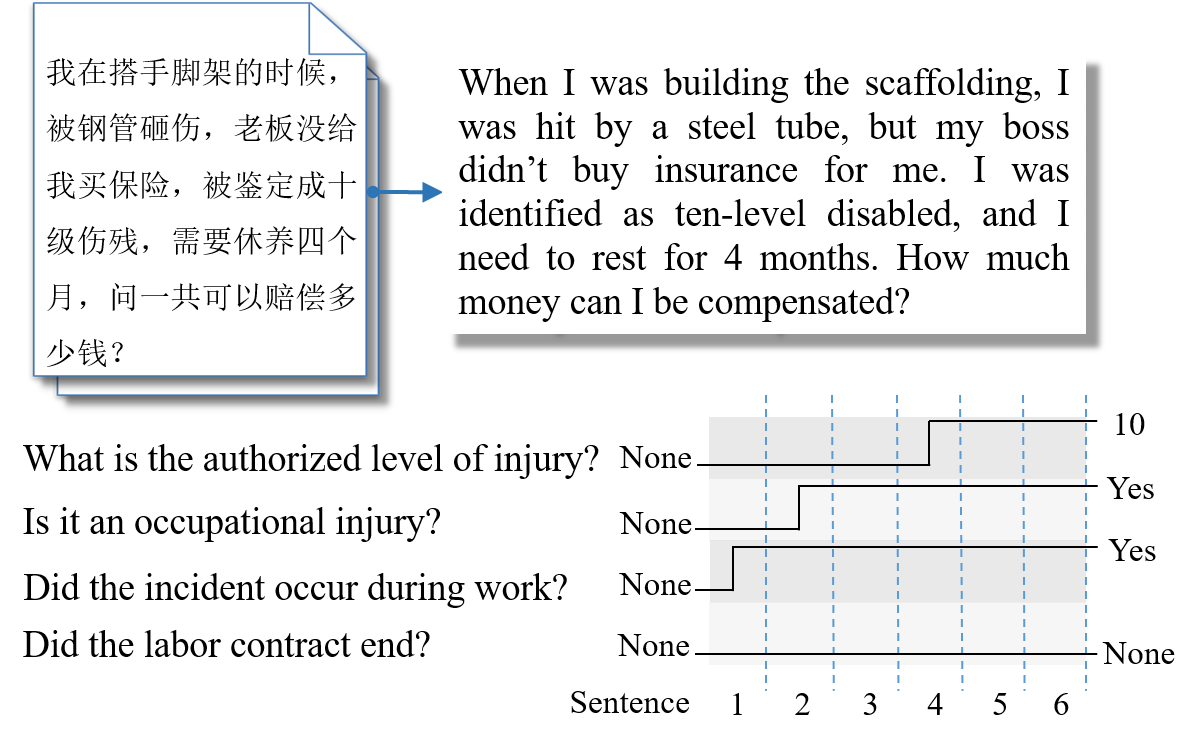}
	\vspace{-5mm}
	\caption{Illustration of \jumper's decision process. Based on a paragraph of six sub-sentences, \jumper\ makes a prediction at an appropriate step for each subtask.}\label{f:intro-task}
		\vspace{-2mm}
\end{figure}

This paper provides a novel framework that models text understanding as a sequential decision process. Our work is inspired by the cognitive process of humans: during reading, people look for clues, perform reasoning, and obtain information from text. We mimic this process by feeding text to a neural network in a sentence-by-sentence manner. At each sentence, the network makes decisions (also known as \textit{actions}) based on the input, and at the end of this process, the network would have some ``understanding'' of the text.

In particular, we focus on text classification problems with several predefined subtasks (called \textit{slots}). When our neural network reads a paragraph, a slot is assumed to have a default value ``\texttt{None}'' at the beginning. At each decision step,  a sentence of the paragraph is fed to the neural network in order; the network then decides if it is confident enough to ``jump'' to a non-default value as the prediction for a particular slot. We impose a constraint that each jump is a finalized decision, which cannot be updated in the future. We call our model \jumper, and its decision process is depicted in Figure~\ref{f:intro-task}.

We train \jumper\ by reinforcement learning with only weak supervision. In human reading, people are typically certain about reading comprehension results, but it is sometimes difficult to model how human belief changes when they read. Likewise, we also assume our training labels only contain the ultimate results, and no supervision signal is given regarding which step the model should make a decision. 

Intriguingly, the one-jump constraint forces our model to be serious about both when to predict and what to predict. This is because a paragraph does not contain a special symbol indicating the end of the paragraph. If our model defers its decision later than it could have made an accurate enough prediction, it takes a risk of not being able to predict. On the other hand, if the model predicts too early, it takes a risk of low accuracy. By optimizing the expected reward in reinforcement learning, the model learns how it can make decisions at a ``right'' time.

The advantage of modeling text classification as a decision process is multi-fold: (1) \jumper\ coincides with recent work on rationalizing neural prediction~\cite{Lei16rationale} when the evidence of classification is local and isolated. (2) In those tasks where information is scattered more widely, \jumper\ learns to make a decision as long as it is confident enough, making it possible to skip reading the remaining part of a paragraph. (3) In a neural model, the evidence of one classification might get distorted after seeing irrelevant facts due to the distributed representation of knowledge. The (partial) decisions that our model has made can serve as valuable ``symbolic'' knowledge.

We evaluated our model on two benchmark datasets; we also collected a new corpus (and make it publicly available) to further evaluate our model in a real, industrial task.
Experiments show that our \jumper\ achieves comparable or higher ultimate classification accuracy compared with strong baselines. Moreover, it reduces the length of text reading by 30--40\%, resulting in fast inference.
For information extraction-style classification where the information is centered in a single sentence, our model can automatically find the key rationale without training signals of jumping positions.
We also show that, in a multitask setting, feeding back the partial decisions (called a \textit{decision-sharing mechanism}) further improves model performance, which indicates that some decisions could help others, serving as symbolic knowledge.

\section{Related Work}

Text classification is related to various tasks in NLP, ranging from sentiment analysis \cite{sentiment} to topic classification~\cite{Wang2012Baselines}. In early years, text classification uses hand-crafted features or feature templates (\eg\ bag-of-words features), based on which machine learning models are used for classification. Recently, deep neural networks have become a prevailing learning model, as they are more powerful classifiers that can work with raw input of words~\cite{Kim14sent}.

Recently, researchers focus more on the rationales underlying neural predictions. \newcite{zhang2016rationale} show that with human annotated rationales, the neural networks' performance could be improved. \newcite{Lei16rationale} build a neural text classifier on key phrases in a paragraph, where key phrase extraction is learned by reinforcement learning with real-valued reward. However, their method cannot deal with non-existing information because it is unclear how to train and predict without extracted phrases. Also, such approach would be more difficult to train with sparse reward (like 0-1 loss).

\newcite{skim} learn to skim text by predicting how many words to skip during reading. However, it is counter-intuitive that a network can learn to skip several future words (which by themselves have a lot of freedom) without actually seeing them. By contrast, our network skims text by ignoring all future sentences after it has been confident enough to predict, where the confidence is said in terms of its expectation of the remaining sentences. 

Different from existing approaches, our paper models text classification as a sequential decision process. The network is similar to the belief tracker in \newcite{bt} for a task-oriented dialog system. However, their network is trained by cross-entropy loss with strong supervision of the groundtruth labels at every step. We instead propose a one-jump constraint in the decision process and train our network by reinforcement learning with weak supervision. 

\input{model}

\input{Experiments} %
\input{Conclusion}


\bibliographystyle{named}
\bibliography{ijcai18}

\end{document}

%% file: model.tex
\section{The Proposed Method}
\label{s:overview}

\begin{figure}[t]
	\centering
	\includegraphics[width=0.47\textwidth]{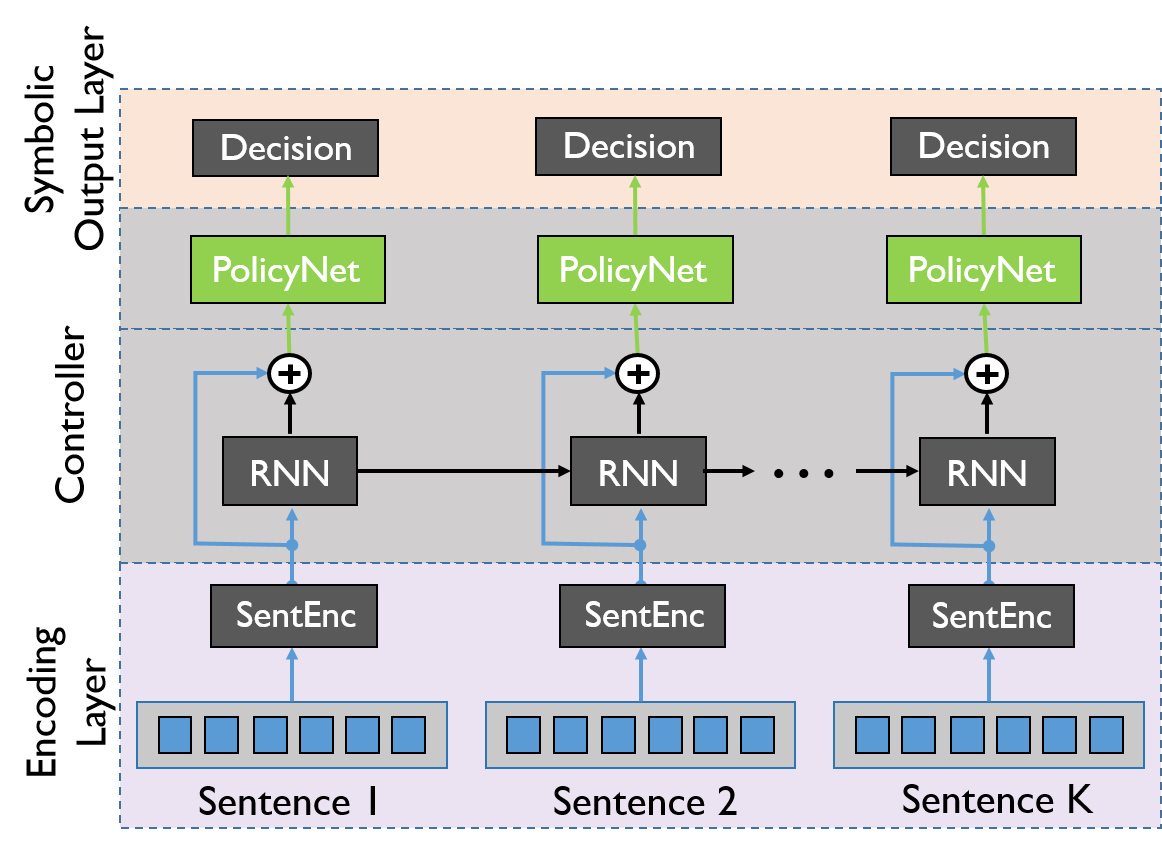}
	\caption{Overview of \jumper. (SentEnc refers to a sentence encoder.)}
	\label{f:overview}
\end{figure}

Figure~\ref{f:overview} shows the overall framework of our approach. We first segment the paragraph into sub-sentences\footnote{Segmented by ``,.!?'' We abuse the terminologies of \textit{sentence} and \textit{sub-sentence} for simplicity if not confusing.}; each could be thought of as a basic unit for some ``proposition,'' and is fed to our model in order. There are three main components in our neural network:
\begin{compactitem}[$\bullet$]
	\item  A sentence encoder encodes the semantic features of words in a sentence into a fixed-dimensional vector space.
	\item  A controller, essentially a recurrent neural network (RNN), is built upon sentence encoders, and takes actions (``jumps'') when appropriate. For each slot, we model it as a classification problem, where a default value ``\texttt{None}'' (indicating information not existing) is included as the classification objective. In other words, the controller decides not only when to jump, but also where (which class) to jump.
	\item A symbolic (output) layer maintains the decisions that have been made, and ensures consistency according to hard constraints that we impose. In this work, we consider a one-jump constraint that allows at most one jump from ``\texttt{None}'' to others. The classification results are the symbolic output layer's values after the network reads the entire paragraph.
\end{compactitem}
\jumper\ is trained by reinforcement learning with weak supervision at the end of a paragraph.
The rest of this section elaborates these components and the training process.

\subsection{Sentence Encoder}
We use a convolutional neural network (CNN)  as the sentence encoder \cite{Kim14sent}. 
CNN applies a set of sliding windows to the concatenation of neighboring words to extract local features, which are aggregated by max pooling to represent sentence-level information. 

For a particular sentence in a paragraph, we denote the word embeddings by $\bm x_1,\bm x_2,\cdots, \bm  x_L\in\mathbb{R}^d$ , where $L$ is the number of words in the sentence and $d$ is the dimension of embeddings. We also denote the concatenation of column vectors   $\bm x_i,  \bm x_{i+1}, \cdots, \bm x_{j}$ by
$\bm x_{i:j} = [\bm x_i \oplus  \bm x_{i+1} \oplus \cdots \oplus\bm x_{j} ]$.

Then convolution is computed by
\begin{align}
c_{k,i} &= f(\bm w_{k}^\top \bm x_{i:i+h-1} + b_k) \\\label{eqn:max}
c_k &= \max\{c_{k,1}, c_{k,2}, \cdots, c_{k,L-h+1} \} \\
\bm c &=  [ c_1 \oplus c_2\oplus \cdots \oplus c_{K} ]
\end{align}
where $\bm{w}_k \in \mathbb{R}^{hd}$ and $b_k\in\mathbb{R}$ are the weights of the $k$th convolutional kernel, extracting a local feature $c_{k,i}$ at position~$i$. The maximum feature over all positions is chosen as the sentence's representation in terms of this kernel. Finally, the features of different kernels are concatenated as the encoding of the sentence, denoted as $\bm c$.

We would like to point out that other networks (\eg\ recurrent neural networks) may also be a reasonable architecture for the sentence encoder. In our work, we choose CNN because we hope to further induce  word-level rationales by backtracking through the max-pooling layer, as will be described in Subsection~\ref{ss:backtrack}.

\subsection{Controller}

Based on encoded sentence features, the controller of \jumper\ takes corresponding actions, as in a sequential decision process. Inside the controller are two submodules: (1) An RNN fuses the current input and previous sentences, maintaining dependency over the entire history; and (2) A policy network (PolicyNet) makes a decision for each slot at the current step. (See also Figure~\ref{f:overview}.)

Formally, RNN takes a sequence of sentence features $\bm{c}_1,\cdots,\bm{c}_T$ and updates its hidden states accordingly. ($T$ is the number of sentences.) In this paper we use the gated recurrent unit (GRU)~\cite{Cho14-GRU} as our recurrent update: $\bm h_t = \gru(\bm h_{t-1},\bm{c}_t)$, where $\bm h_t$ is the hidden state of the time step~$t$.

Based on RNN's hidden states, PolicyNet predicts the decision action for each slot. Suppose slot $i$ has $N_i$ possible values, we use a softmax predictor of $N_i+1$ ways, where an additional way ``\texttt{None}'' represents information not existing. Notice that the ``\texttt{None}'' class does not differ from other classification labels at the beginning of training. However, the reinforcement learning with the one-jump constraint would make the model predict ``\texttt{None}'' before it is confident enough to take an action.

Formally, \jumper's decision $\bm a_t^{(i)}\in\mathbb{R}^{N_i+1}$, after processing the $t$th sentence, is given by a policy distribution
\begin{align}
 \pi(\bm a_t^{(i)}|\bm c_t) &= \operatorname{softmax}(W_{p}^{(i)}[\boldsymbol{c}_t \oplus\bm h_{t}]+\bm b_p^{(i)})
\end{align}
where $W_p$ and $\bm{b}_p$ are weights and the bias term. Here, we feed PolicyNet with the concatenation of RNN's hidden state and sentence features, inspired by ResNet~\cite{He2016resnet}.
During training, we sample an action from its predicted distribution, whereas for testing, we choose the action with the maximum \textit{a posteriori} probability, \ie\ $a_t^{(i)}=\operatorname{argmax}\pi(a_t^{(i)}|\bm c_t)$.

PolicyNet appears to resemble a multitask sequential labeler for different slots based on a shared hidden representation. This makes sense because different slots may be correlated with each other (\eg\ the occupational injury task in Table~\ref{f:intro-task}), and thus some underlying representations could be reused.

However, PolicyNet cannot be trained with standard cross-entropy loss due to the lack of step-by-step supervision. Instead, we apply reinforcement learning, which computes the policy gradient to optimize the expected reward (described in Subsection~\ref{ss:learn}). In this way, PolicyNet learns in a trial-and-error fashion.

\subsection{Symbolic Output Layer}

The output layer of \jumper\ keeps the decisions that have been made until the current sentence, and maintains consistency in a ``symbolic'' fashion. We propose a one-jump constraint that allows at most one non-default prediction (not ``\texttt{None}'') for each slot. 

Let $\bm s_t^{(i)}\in\{0,1\}^{N_i+1}$ be the one-hot representation of the symbolic layer's state for slot $i$ at the $t$th sentence. We have
\begin{align}
\bm s_t^{(i)} &=\bm s_{t-1}^{(i)}\cdot \boldsymbol{1}_{\{s_{t-1}^{(i)}\ne\texttt{None}\}} +\bm a_t\cdot \boldsymbol{1}_{\{s_{t-1}^{(i)}=\texttt{None}\}}
\end{align}
where $ \boldsymbol{1}_{\{\cdot\}}  $ is an indicator function that yields 1 when its argument is true, and 0 otherwise. In other words, the state has to be the same as its previous one if it is not ``\texttt{None}''; but on the other hand, \jumper\ can remain in ``\texttt{None}'' for the entire paragraph if the information does not exist.

The final result, which we use to compare with groundtruth, is the symbolic layer's state after processing the last ($T$th) sentence, \ie\  $s_T^{(i)}$.

It should be mentioned that, although our current \jumper\ considers only one constraint, it is natural to design other reasoning rules within the symbolic layer, like the way in OONP\cite{oonp2018}, for example, one slot inferring another. Future work is needed to address more complicated symbolic reasoning in the \jumper\ framework.

\subsubsection{Decision-Sharing Mechanism}

In a multi-slot prediction task, some information may be useful for others. Thus we feed the symbolic states back to the neural network, so that different action predictors are aware of each other. We call this a \textit{decision-sharing mechanism}. Concretely, we concatenate one-hot representation of each state with the sentence vector and feed them to $\gru$, given by
$\bm h_t = \gru(\bm h_{t-1},[c_t\oplus \bm s_{t-1}^{(1)}\oplus \cdots\oplus\bm s_{t-1}^{(N_s)}])$,
where $N_s$ is the number of slots.


As we shall see in experiments, the decision-sharing mechanism helps to improve the accuracy of both rationale finding and ultimate classification.

\subsection{Learning}\label{ss:learn}
The main difficulty of learning \jumper\ is the lack of step-by-step supervision, \ie\ we assume the labels contain only ultimate results for each slot, but no information for the appropriate position to jump. Admittedly, life will be easier if we have fine-grained annotations regarding which step to predict for each slot, but they are costly and labor-intensive to obtain. 

We therefore apply reinforcement learning to train our \jumper\ framework. We define a reward by comparing the model's prediction and the groundtruth. The training objective is to maximize the expected reward over sampled actions.

Concretely, we define the final reward of $\jumper$ as
\begin{equation}
R_\text{final}^{(j,i)}=\bm 1_{\{\bm s^{(j,i)}_{T_j} = \bm t^{(j,i)}\} }
\end{equation} 
for data point $j$ and slot $i$, where $\bm s^{(j,i)}_{T_j}$ is the symbolic state at the end of the paragraph and $\bm t^{(j,i)}$ is the groundtruth.

However, it is difficult to train a model by reinforcement learning, with such a sparse reward along the decision process. In particular, our model tends to jump at early stages at the beginning of training, because for an uniform distribution over all $N+1$ actions, the probability of jumping at time step~$t$ is $\frac N{(N+1)^t}$. Thus we design an intermediate reward as
\begin{equation}
R_\text{int}^{(j, i,t)}=\left\{
\begin{aligned}
r,  \;\;\;&\text{if $ s_t = \texttt{None}$ } \\
0,   \;\;\;&\text{otherwise}\\ 
\end{aligned}
\right.
\label{e:reward_zero}
\end{equation} 
for data step $j$, slot $i$, and each step $t$. Here, $r$ is a (small) positive constant, balancing the importance of $R_\text{int}^{(\cdot)}$ and $R_\text{final}^{(\cdot)}$. We would like to emphasize that the design of $R_\text{int}^{(\cdot)}$ is different from traditional planning and reinforcement learning (\eg\ the maze problem) where the reward for each step is negative. In our problem, however, each step has a positive reward so that it alleviates the early-jumping problem.

We compute the cumulated reward from step $t$ to the jumping step as:
\begin{equation}
R_{t:T_\text{jump}^{(j,i)}}^{(j,i)} = \sum_{t'=t}^{T_\text{jump}^{(j,i)}} \gamma^{t'-t}R_\text{int}^{(j,i,t')} + R_\text{final}^{(j,i)}  \vspace{-1mm}
\end{equation} 
where $\gamma$ is the discounting rate, and $T_\text{jump}^{(j,i)}$ denotes the jumping step.  To maximize the expected reward, we compute the gradient of the policy, given by
\begin{align}\nonumber \vspace{-2mm}
\nabla_\Theta\mathcal{J}\!(\Theta)\!=\!\mathbb{E}_{\pi_\Theta}\!\!\left[\sum_{t=1}^T\nabla_\Theta R_{t:T}\log \pi_\Theta\!\!\left(\bm a_t | \bm c_t \right) \right] \\  \nonumber
\approx\! 
\sum_{j=1}^N\sum_{i=1}^I\sum_{t=1}^{ T_\text{jump}^{(j,i)}}\frac{1}{NT_j} R_{t:T_\text{jump}^{(j,i)}}^{(j,i)}\nabla_\Theta\log \pi_\Theta\! \left(\bm a_t^{(j,i)} |\bm c_t^{(j)} \right) 
\end{align}
where $\Theta$ denotes all model parameters and $I$ denotes the number of slots. The approximation is due to Monte Carlo sampling 
and the above updating rule is also known as the REINFORCE algorithm~\cite{williams92-reinforce}.  To have a balance between exploration and exploitation, we reserve a small probability $\epsilon$ to uniformly sample from the entire action space.
To reduce the variance of REINFORCE, we subtract the reward by a baseline term (computed as the average of the $M=5$ samples) and truncate negative rewards as in~\newcite{coupling}.

It is interesting to have an intuitive understanding on why \jumper\ can find the ``right'' position to predict with only weak supervision. Let $t_*$ be the position that the network could have predicted. The reward encourages the model to predict any time after $t_*$, and later sentences have a slightly higher reward due to $R_\text{int}$. However, if the network learns to predict as late as possible by maximizing the reward for a particular training data point, it has to wait for intermediate reward by not predicting. Since there is no clue indicating the end of a paragraph, the network unfortunately cannot learn such information, and thus is in the risk of not being able to predict for other samples, resulting in a low total reward over the training set.
Therefore, our \jumper\ framework with the one-jump constraint enables the model to find the ``right'' position to predict by weak supervision.

\subsection{Backtracking Word-Level Clues}
\label{ss:backtrack}

Currently, our approach works in the sentence level. To obtain word-level rationales in our model, we propose a simple heuristic that backtracks information flow through max-pooling operation, based on the key sentence that we have already found by \jumper.
We compute the gradient of the log-likelihood with respect to the last sentence's representation; it is then multiplied with the magnitude of the difference between two steps (ignoring the sign by taking the square). The two aspects indicate how a feature (at the last step) could have improved the prediction, and what is mostly changed at the current time step. Then we choose the top $D=10$ values, yielding the most important $D$-dimensions in the output of CNN, given by 
\begin{equation}
\mathscr{D} =  \operatorname{top}_D\left( \frac{ \partial \log(p_t(s_t))}{\partial \boldsymbol{c}^{(t-1)}} \odot ( \boldsymbol{c}^{(t)} - \boldsymbol{c}^{(t-1)})^2\right) 
\end{equation}
where $\odot$ indicates point-wise product.

We backtrack where the maximum values come from in the max-pooling operation in Equation~\ref{eqn:max}, obtaining the word that matters in a dimension $d$ as $w_d = \operatorname{argmax}\{c_1, \cdots, c_K\}$.
The importance of a word is counted as the fraction in $\mathscr D$ at which the word is backtracked.

%

%% file: Experiments.tex
\section{Experiments}
We evaluated \jumper\ on three tasks, including two benchmark datasets and one real, industrial application. We show the performance of both ultimate classification and jumping positions; we will also have deep analysis into our model.

\begin{table}[!t]
	\center
	\tiny
	\resizebox{0.8\linewidth}{!}{
		\begin{tabular}{|c|c|c|c|c|c|c|}
			\hline
			\textbf{Data} & \# of class & \# of samples &\# of vocab  & {Test} \\ \hline
			MR & 2  & 10,662 &  18,765 & 10-fold \\ 
			AG &  4 &  127,600 &  17,836  & 7,600 \\ 
			OI &  2--12  &  3,995 & 2,089 & 400  \\
			\hline
		\end{tabular}
	}\vspace{-1mm}
	\caption{Statistics of the datasets after tokenization: the numbers of classes, data samples, vocabulary size, and test samples. For MR which does not have a standard split, we performed 10-fold cross-validation.}
	\label{tab:data-stat}\vspace{-2mm}
\end{table}

\subsection{Datasets}  \label{s:dataset}

In this part, we describe the datasets used in our experiments.
\begin{compactitem}[$\bullet$]
	\item Movie Review (MR), whose objective is a binary sentiment classification (positive vs.~negative) for movie reviews~\cite{MR}; it is widely used as a sentence classification task.
	\item  AG news corpus (AG), which is a collection of more than one million news articles, and we followed \newcite{Zhang15-cnnText}, classifying the largest four categories: \texttt{world}, \texttt{sports}, \texttt{business}, and \texttt{science}.
	\item Occupational Injury (OI).\footnote{Both code and the Occupational Injury dataset are available at: https://github.com/jumper-data}
	The task---information extraction of occupational injury---originates from a real industrial application in the legal domain. 
	We constructed a dataset (in the Chinese language) of 3995 cases related to occupational injuries from an online domain-specific forum. Based on an established ontology with 15 slots, each text is annotated with answers for these 15 problems. Table \ref{tab:dataset2} shows some statistics of the OI dataset. We report two subtasks---occupational injury identification (InjIdn) and injury level (Level)---to evaluate our model in a single-task setting. We used all subtasks to evaluate the decision-sharing mechanism.
\end{compactitem}

\begin{table}[!t]
	\centering
	\tiny
	\resizebox{0.75\linewidth}{!}{
		\begin{tabular}{|l|c|c|c|c|}
			\hline
			Subtask & \# of Class & Majority Guess (\%)  \\
			\hline
			IsOccuInj &  2 & 85.68 \\
			AssoPay  & 2 & 81.75 \\
			LaborContr & 2  &  93.22 \\
			EndLabor & 3  &  93.67 \\
			OnOff & 2  &  93.69 \\
			DiseRel & 3  &  99.05 \\
			OutForPub & 2  &  99.07 \\
			WorkTime & 3  &  79.75 \\
			WorkPlace & 3  &  80.60 \\
			JobRel & 3  &  91.34 \\
			\textbf{InjIdn} & 3  &  55.02 \\
			ConfirmLevel & 3  &  72.99 \\
			Insurance & 3  &  89.66 \\
			HaveMedicalFee & 3  &  83.63 \\
			\textbf{Level} & 12  &  82.65 \\
			\hline
		\end{tabular}
	 }\vspace{-1mm}
		\caption{Statistics of the Occupational Injury (OI) dataset. We chose injury identification (InjIdn) and injury level (Level) as the tasks for single-slot prediction, highlighted in bold; all were used to evaluate the decision-sharing mechanism. Details of the dataset can be found in Footnote 2.}
		\label{tab:dataset2}\vspace{-2mm}
	\end{table}
	
	\subsection{Competing Methods}\label{ss:baseline}
	We compare \jumper\ with the following baselines:
	\begin{compactitem}[$\bullet$]
		\item \textbf{Hierarchical CNN-GRU}. We use \jumper\ with cross-entropy loss as a baseline model, which is essentially a Hierarchical model with CNN and GRU for sentences and paragraphs, respectively. This baseline is similar to our model except training criteria.
		\item \textbf{Bi-GRU}. It reads a text in two opposite directions, and the final states are concatenated for prediction. 
		\item \textbf{CNN}. This model is proposed by~\newcite{Kim14sent}, with several different sizes of convolution operators to learn sentence representation.
		\item \textbf{Self-Attentive}. \newcite{selfatt-lin2017} propose a self-attentive model that attends to the sequence itself.
	\end{compactitem}
	In the latter three baselines, we concatenated all sentences, and the models were applied to the paragraph.
	\subsection{Implementation Details}
	In our experiments, we applied coarse grid search on both the MR and OI development datasets to select hyperparameters. We did not perform any dataset-specific tuning except early stopping on the development sets. For AG, which does not have a standard split, we randomly selected 5\% of the training data as the development set.
	
	In our model and baselines, the CNN part used rectified linear units (ReLU) as the activation function, filter windows with sizes 1 to 5, 200 feature maps for each filter, and a dropout rate of 0.5; GRU had a hidden size of 20. We re-implemented the self-attentive model using the same hyperparameters as in \newcite{selfatt-lin2017}.
	
	For reinforcement learning, the intermediate reward $r$ was 0.05, discounting rate $\gamma$ was 0.9, and the exploration rate $\epsilon$ was 0.1.
	
	In addition, word embeddings for all of the models were initialized with 300d GloVe vectors~\cite{pennington2014glove} and fine-tuned during training to improve the performance. The other parameters were initialized by randomly sampling from the uniform distribution in $ [-0.01, 0.01]$.
	For all the models, we used AdaDelta  with a learning rate of 0.1 and a batch size of 50.
	
	\subsection{Results and Discussion}
		In this section, we present \jumper's performance regarding several aspects: overall accuracy, jumping accuracy, and multi-slot learning. We also present a case study to showcase the behavior of our model. 
	
	\textbf{Classification Results.} We first analyze the classification accuracy of \jumper\ when compared with baselines. Table~\ref{tab:perf-res} shows the test performance on the three datasets with four tasks. We notice that, in the MR and AG datasets, \jumper\ occasionally predicts  ``\texttt{None},'' which is not a valid label in these datasets. This puts our model at a disadvantage, and we take the most likely non-default (not ``\texttt{None}'') as the prediction at the end of a paragraph. 
	
	As shown, our \jumper\ model achieves comparable or better performance on all these tasks. This indicates that modeling text classification as a sequential decision process does not hurt or even improves performance. We would also like to point out that ``accuracy" is not the only performance that we are considering. More importantly, our proposed model is able to find the key supporting sentence for text classification, or reduce the reading process, as shown in the following experiments.
\begin{table}[t]
	\center
	\resizebox{0.9\linewidth}{!}{
		\begin{tabular}{lccccccc}
			\toprule
			\!\!\!\textbf{Model} && \makebox[1.cm]{\textbf{MR}} & \makebox[1.cm]{\textbf{AG}} & \makebox[1.cm]{\!\!\!\textbf{OI-Level}} & \makebox[1.cm]{\textbf{OI-InjIdn}}  \\ 
			\midrule
			\!\!\!CNN$^\dag$~\cite{Kim14sent} &&  81.00 & -- & -- & -- \\ 
			\!\!\!fasttext$^\dag$~\cite{Joulin16Fasttext}\hspace{-.7cm} &&  -- & 92.50 & -- & -- \\  
			\midrule
			\!\!\!Bi-GRU && 77.80  & 92.44   &  94.75  & 73.25  \\ 
			\!\!\!CNN &&  80.80 & 92.58 &  96.25  & 74.25   \\ 
			\!\!\!Self-Attentive &&  \textbf{82.10} & 91.40 & 97.00 & 73.25 \\ 
			\!\!\!Hierarchical CNN-GRU && 80.23 & 92.49 & 95.75 & 74.75  \\
			\midrule
			\!\!\!\jumper			  &&  80.67  & \textbf{92.62} & \textbf{97.25}  & \textbf{75.50}  \\
			\bottomrule
		\end{tabular}
	}\vspace{-1mm}
	\caption{Test accuracy (\%) on MR, AG, and OI datasets. $^\dag$Results quoted from previous papers.}
	\label{tab:perf-res}
			\vspace{-2mm}	
\end{table}

\begin{table}[!t]	
	\centering
	\resizebox{.85\linewidth}{!}{
		\begin{tabular}{lcccc}
			\hline
			\!\!\!Dataset & MR & AG &\!\!\!\!\! OI-Level\!\! &\!\!\!\!\! OI-InjIdn\!\!\!\\
			\hline
			\!\!\!Avg \# of sub-sentences\!\!   & 2.17   & 3.46 & 4.88  & 4.88  \\
			\!\!\!Avg jumping position\!\!  & 1.46   & 2.04 & 3.23  & 2.87  \\	
			\!\!\!Reduced \%  \!\!    &\!\! 32.7\%\!\! &\!\! 41.0\%\!\!&\!\!33.8\%\!\! &\!\! 41.2\%\!\!\\
			\hline
		\end{tabular}
	}\vspace{-1mm}
	\caption{Statistics of the average number of sub-sentences, the average jumping position and the proportion of reduced text. The one-jump constraint enables the model to skip future sentences after a decision is made.}
	\label{tab:skip}
			\vspace{-2mm}
\end{table}
\begin{table}[t] 
	\centering
	\resizebox{.7\linewidth}{!}{
		\begin{tabular}{lccc}	
			\toprule	
			{\textbf{Model}} &  \textbf{CA}  & \textbf{JA} &  \textbf{OA} \\ 
			\midrule	
			CNN & 96.25 & 94.81 &  91.25  \\
			Self-Attentive & 97.00 & 98.45 &  95.50  \\
			Hierarchical CNN-RNN\quad& {96.00}  &  {98.18} & {94.25}\\ 
			\jumper  & \textbf{97.25} &  \textbf{100}& \textbf{97.25} \\  	
			\bottomrule\end{tabular}
	}\vspace{-1mm}
	\caption{Performance of finding the key rationale in the OI-Level dataset, where information is often local. \textbf{CA}: Classification accuracy. \textbf{JA}: Jumping accuracy. \textbf{OA}: Overall accuracy.}\vspace{-3mm}
	\label{tab:rationale}	
\end{table}

	\textbf{Performance of Jumping.} \jumper\ has to make a decision as long as it sees sufficient evidence during its reading process due to the one-jump constraint, and after prediction, there is no need to read future sentences. We see in Table~\ref{tab:skip} that, although our model achieves similar or higher performance compared with strong baselines, it reduces the length of text reading by 30--40\%, leading to fast inference for prediction.
	
	We are now further curious if \jumper\ could ``jump'' at the right position in an information extraction-style task such as OI-Level. We annotate the rationale sentences in 400 data points (also available on our website in Footnote~2), serving as the test groundtruth. It should be noticed that we still have no training labels for jumping positions in this experiment. We compare \jumper\ with the Hierarchical CNN-GRU model, which uses the same neural network, but differs in terms of training methods. The Hierarchical CNN-GRU is trained with cross-entropy loss at the end of a paragraph. During testing, we apply the predictor to every step and find the first position that it makes a prediction other than ``\texttt{None}.'' This heuristic makes some sense, because the RNN is supposed to map information to the same hidden space during its recurrent modeling of a sequence. 
	We also included a competing method that uses a CNN classifier~\cite{Kim14sent} and chooses the sentence where words are selected the most by max pooling.

 In addition to the classification accuracy (CA) shown in Table~\ref{tab:perf-res}, we use the following metrics: (1) Jumping accuracy (JA), the percentage of correct jump positions conditioned on correct classification; and (2) Overall accuracy (OA), the percentage of both correct jumping positions and correct classification results. We also include the classification accuracy (CA) as has been shown in Table~\ref{tab:perf-res}. It is easy to verify that $\text{OA}=\text{CA}\cdot\text{JA}$. 
	
		The results are shown in Table~\ref{tab:rationale}. We see that \jumper\ can discover the jumping position with a very high accuracy in terms of both JA and OA, and that both CNN and Hierarchical CNN-GRU perform worse in this task. Although they achieve similar classification results (\jumper\ slightly outperforming by $\sim$1\%), \jumper\ is better at finding the key rationale by 3--6\%. This shows that our one-jump constraint forces the model to think more carefully about when to make a decision, and that reinforcement learning is an effective way to learn the correct position of making decisions.

Another interesting finding is that, for Hierarchical CNN-GRU, the classification accuracy at the end of the paragraph as in Table~\ref{tab:perf-res} is lower than that when it could have predicted as in Table~\ref{tab:rationale}. This shows evidence of the distortion phenomenon of distributed representation: when neural networks are fed with too much irrelevant information, its knowledge is less accurate. 

\textbf{Evaluating the Decision-Sharing Mechanism.}
	We now evaluate \jumper\ in a multitask learning setting to see if the symbolic knowledge can help decision making for other slots. The average accuracy and $F_1$ scores for the 15 OI subtasks are shown in Table~\ref{tab:multitask-exp}. We include $F_1$-score because some slots are skewed. We see that \jumper\ achieves better performance, with known knowledge formatted in a symbolic way and fed back to the neural network. Although the improvement is not large, the results are consistent in terms of both accuracy and the $F_1$-score for both development and test sets.

	\begin{table}[!t]
		\centering
		\resizebox{.8\linewidth}{!}{
		\begin{tabular}{|c|c|c|c|c|}
			\hline
			\multirow{2}{*}{\textbf{Model}} & \multicolumn{2}{c|}{\textbf{Accuracy}} & \multicolumn{2}{c|}{$F_1$} \\ \cline{2-5} 
			& dev           & test          & dev        & test       \\ \hline
			Bi-GRU                  & 90.62         & 90.18         & 20.02      & 20.20      \\ 
			CNN                     & 92.41         & 91.64         & \textbf{30.99}      & 29.05      \\ 
			Self-Attentive          & 92.12         & 91.85        & 21.26      & 22.61      \\ 
			Hierarchical CNN-GRU    & 91.56         & 91.30         & 24.80      & 24.44      \\ 
			\jumper                 & 92.43         & 92.42         & 26.57      & 29.60      \\ 
			\jumper-sharing         & \textbf{92.71}         & \textbf{92.65}         & 27.57      & \textbf{30.52}      \\ \hline
		\end{tabular}
		}\vspace{-1mm}
		\caption{The average accuracy and $F_1$ on the OI dataset using the decision-sharing mechanism.}
		\label{tab:multitask-exp}
		\vspace{-3mm}
	\end{table}
	
\begin{figure}[!tbp]
		\centering
		\includegraphics[width=0.5\textwidth]{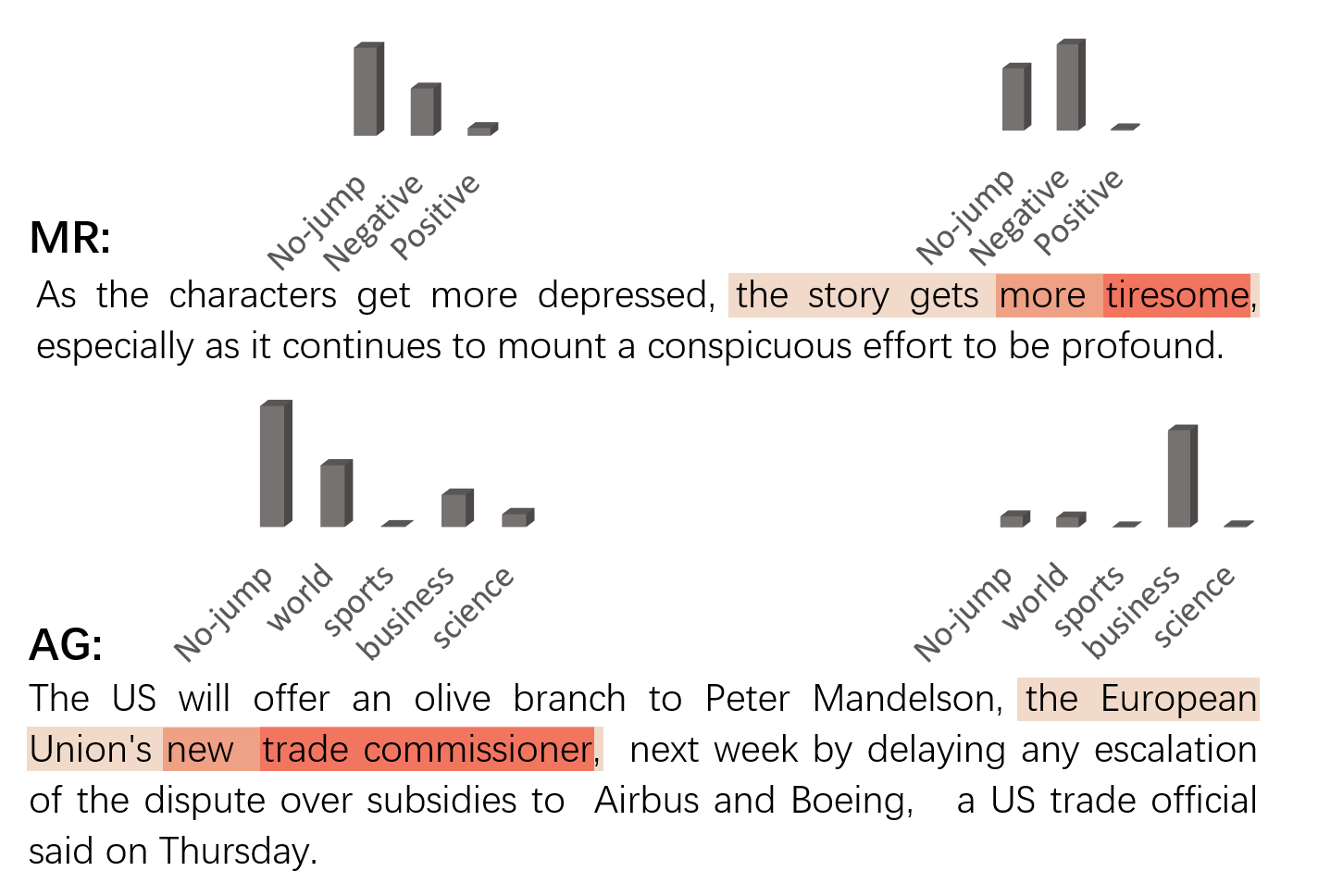}
				\vspace{-6mm}
		\caption{Case study. We show the histogram of decision distributions and the heatmaps of word importance in MR and AG samples. }
		\label{f:showcase}
						
	\end{figure}

	\textbf{Case Study.}
	We show several examples of the decisions made by the neural network in Figure~\ref{f:showcase}. In the AG and MR datasets, information is  located over a wider range, and the network makes a prediction as long as it sees enough evidence (\eg\ ``trade commissioner'' for the business domain). By backtracking the word-level rationales, we find words like  ``trade commissioner'' and ``tiresome''  play a more important role in the decision making. In these cases, the model does not need to read future sentences, which is more efficient than reading the entire paragraph.
	For OI-Level classification where information is mostly local, the neural network precisely locates the subsentence that contains the information, as shown in Table~\ref{tab:rationale}. 
		\vspace{-1mm}

%% file: Conclusion.tex
\section{Conclusion and Future Work}
In this paper, we have proposed a novel model, \jumper, that models  text classification as a sequential decision process on a sentence-by-sentence basis when reading a paragraph. We train \jumper\ by reinforcement learning with a one-jump constraint. Experiments show that \jumper\ achieves comparable or higher performance than baselines; that it reduces text reading by a large extent; and that it can find the key rationale if the information is local within a sentence.

In future work, we would like to incorporate symbolic reasoning into the symbolic output layer, where we could explicitly handle inference, contradiction, etc.~among different slots.

\section*{Acknowledgments}
We thank anonymous reviewers for their constructive comments. We also thank Daqi Zheng and Fangzhou Liao for their insightful discussion. This work was supported by the  Beijing Innovation Center for Future Chip.